\documentclass[aps,preprint,prd,nofootinbib]{revtex4}
\usepackage{graphicx}
\usepackage{psfrag}

\usepackage{subfigure}
\usepackage{array}
\usepackage{graphicx}
\usepackage{amsmath}
\usepackage{amssymb}
\usepackage{mathrsfs}
\usepackage{bm}
\usepackage{slashed}
\usepackage{threeparttable}
\usepackage{multirow}
\usepackage[colorlinks, linkcolor=black, anchorcolor=black, citecolor=black]{hyperref}
\usepackage[amsmath,thmmarks,hyperref]{ntheorem}
\usepackage{soul}

\allowdisplaybreaks[4]

\begin{document}

\title{Strong decays of $2^+$ charm and charm-strange mesons}
\author{Si-Cheng Zhang\footnote{sichengzhangcn@gmail.com}, Tianhong Wang\footnote{thwang@hit.edu.cn}, Yue Jiang\footnote{jiangure@hit.edu.cn}, Qiang Li\footnote{lrhit@qq.com}, \\and Guo-Li Wang\footnote{gl\_wang@hit.edu.cn}\\}
\address{Department of Physics, Harbin Institute of Technology, Harbin, 150001, China}

\baselineskip=20pt

\begin{abstract}

In this paper, we calculate the strong decays of $2^+$ heavy-light states, namely, the charmed $D^*_2(2460)^0$ meson and the charm-strange $D^*_{s2}(2573)^+$ meson. The method we adopt is the reduction formula, PCAC relation and low energy theorem, following which, the transition amplitudes are calculated. The wave functions of the heavy mesons involved are achieved by solving the instantaneous Bethe-Salpeter equation. As the OZI-allowed two-body strong decays give the dominant contribution, they can be used to estimate to total widths of mesons. Our results are: $\Gamma[D^*_2(2460)^0]=51.3$ MeV and $\Gamma[D^*_{s2}(2573)^+]=19.6$ MeV. The ratios of branching ratios of two main channels are $Br[D^*_2(2460)^0\rightarrow D^+\pi^-]/Br[D^*_2(2460)^0\rightarrow D^{\ast+}\pi^-]=2.13$ and $Br[D^*_{s2}(2573)^+\rightarrow D^{\ast 0} K^+]/Br[D^*_{s2}(2573)^+\rightarrow D^0K^+]=0.08$, respectively.

\end{abstract}

\maketitle

\section{Introduction}

In recent years, many new charm and charm-strange states have been found, which attracted lots of attention. Most of these particles, unlike the ground states $D^{(\ast)}$ and $D_s^{(\ast)}$ whose mass and branching ratios of different decay channels are detected more precisely, still need more careful studies both theoretically and experimentally.  Among these states, the $P$-wave ones are very interesting. On the one hand, almost all of them have been identified and the experimental data are relatively abundant; on the other hand, these states provide a good test ground for the  different phenomenological models.

In this paper, we focus on the OZI-allowed two-body strong decays of $2^+$ charm and charm-strange states, namely, $D_2^\ast(2460)$ and $D_{s2}^\ast(2573)$. Although both particles are found years ago~\cite{cleo, tps}, there is still lack of experimental data for their branching ratios. So theoretical predictions for the partial widths of different channels are important. Of the existing approaches to study two-body strong decays of heavy-light mesons, most applied the simple harmonic oscillation (SHO) wave functions for the mesons involved. For $P$-wave states, as they are the orbital excitation, the relativistic corrections will be considerable, and adopting more reliable wave functions is necessary. Also, the results for these decays given by different models vary a lot. So more careful studies are still needed.

The method applied here consists of two ingredients. First, the transition matrix element is reduced by using PCAC and low energy theorem. Second, the wave functions of initial and final heavy mesons are achieved by solving the instantaneous Bethe-Salpeter equation~\cite{BS1,BS2}. This method has been used in Refs.~\cite{zhang, wzh01} and Ref.~\cite{wth16} to deal with the two-body strong decays of $S$-wave and $D$-wave heavy-light mesons, respectively. There reasonable results were obtained. As pointed in Ref.~\cite{wth16}, the chiral quark model~\cite{zz} also got a similar form of the transition amplitude, while there the SHO wave function was used. So here by applying the Salpeter wave function, the relativistic correction (at least parts of which) can be taken properly.

The paper is organized as follows. In section II, we give the hadronic matrix elements and the formulae for the decay widths of different two-body strong decay channels. In section III, the parameter values involved in the numerical calculation are given. And we show our results and make some discussions. The conclusion is presented in Section IV.

\section{Matirx elements and decay widths}

We take the decay channel $D^{*0}_2\to D^{(\ast)+}\pi^-$ as an example (for $D_{s2}^\ast$, the formulae are the same), whose Feynman diagram is given in Figure 1.
By applying the reduction formula, the transition matrix element can be written as~\cite{chang}
\begin{equation}
\begin{aligned}
       \langle D^{(\ast)+}(P_1)\pi^-(P_2)|D^{*0}_2(P)\rangle=\int\textrm{d}^4xe^{iP_2\cdot x}(M_\pi^2-P_2^2) \langle D^{(\ast)+}(P_1)|\Phi_\pi(x)|D^{*0}_2(P)\rangle,
\end{aligned}
\end{equation}
where $P$, $P_1$, and $P_2$ are momenta of $D_2^{\ast0}$, $D^{(\ast)+}$, and $\pi^-$, respectively. $\Phi_\pi(x)$ is the field of $\pi^-$, which, by using PCAC relation, is expressed as the divergence of a axial vector current
\begin{equation}
      \Phi_\pi(x)=\frac{1}{M^2_\pi f_\pi}\partial^\mu(d\gamma_\mu\gamma_5\bar{u}),
\end{equation}
where $f_\pi$ is the decay constant of $\pi^-$.

Inserting Eq.~(2) into Eq.~(1), we get
\begin{equation}
\begin{aligned}
      \langle D^{(\ast)+}(P_1)\pi^-(P_2)|D^{*0}_2(P)\rangle &=\frac{(M^2_\pi-P_2^2)}{M_\pi^2f_\pi}\int\textrm{d}^4xe^{iP_2\cdot x}\langle D^{(\ast)+}(P_1)|\partial^\mu(d\gamma_\mu\gamma_5\bar{u})|D^{*0}_2(P)\rangle \\
      &=\frac{-\textrm{i}P_2^\mu(M^2_\pi-P_2^2)}{M_\pi^2f_\pi}\int\textrm{d}^4xe^{iP_2\cdot x}\langle D^{(\ast)+}(P_1)|d\gamma_\mu\gamma_5\bar{u}|D^{*0}_2(P)\rangle  \\
      &\approx\frac{-\textrm{i}P^\mu_2}{f_\pi}\int\textrm{d}^4xe^{iP_2\cdot x}\langle D^{(\ast)+}(P_1)|d\gamma_\mu\gamma_5\bar{u}|D^{*0}_2(P)\rangle  \\
      &=(2\pi)^4\delta^{(4)}(P-P_1-P_2)\frac{-\textrm{i}P^\mu_2}{f_\pi}\langle D^{(\ast)+}(P_1)|d\gamma_\mu\gamma_5\bar{u}|D^{*0}_2(P)\rangle,
\end{aligned}
\end{equation}
where in the second equation we used the partial integral, and the third equation is deduced by using the low energy thorem.
Now the strong decay amplitude is represented by Figure 2. The decay amplitude is
\begin{equation}
      T(D^{*0}_2\to D^{(\ast)+}\pi^-)=\frac{-\textrm{i}P^\mu_{f_2}}{f_{\pi}}\langle D^{(\ast)+}(P_1)|d\gamma_\mu\gamma_5\bar{u}|D^{*0}_2(P)\rangle,
\end{equation}
where the contribution of $\pi^-$ is reduced to the factor $\frac{P^\mu_{f_2}}{f_{\pi}}$.

\begin{figure}[ht]\label{Feyn1}
\centering
\includegraphics[scale=1.0]{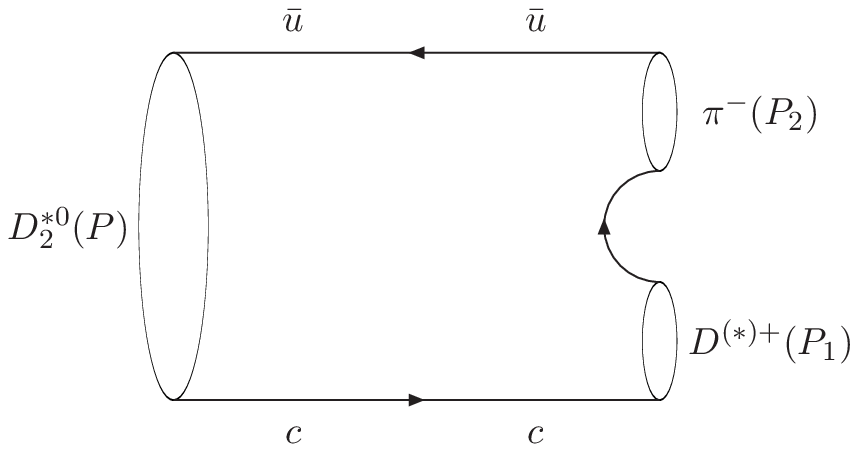}
\caption[]{Feynman diagram for $D^{*0}_2\to D^{(\ast)+}\pi^-$.}
\end{figure}

\begin{figure}[ht]\label{Feyn2}
\centering
\includegraphics[scale=1.0]{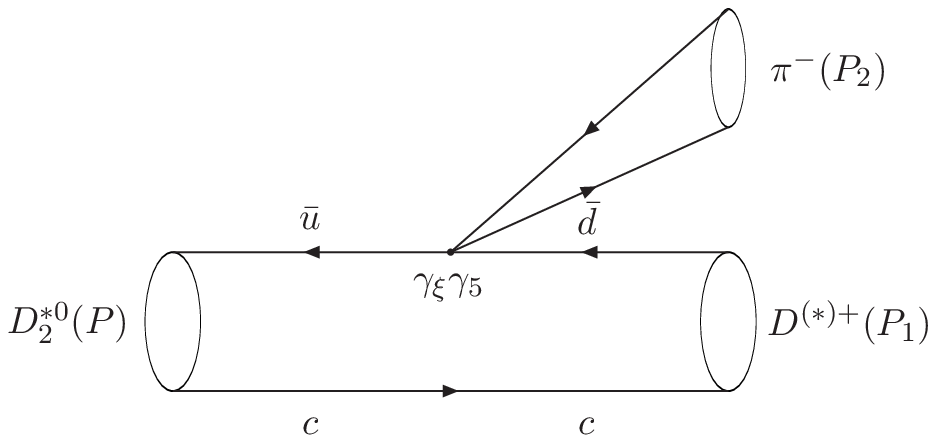}
\caption[]{Feynman diagram for $D^{*0}_2\to D^{(\ast)+}\pi^-$(low energy approximation).}
\end{figure}

As for the decay $D^{*0}_2\to D^0\eta$, the $\eta-\eta'$ mixing should be considered. At this case, the field of $\eta$ is expressed as
\begin{equation}
   \Phi_\eta(x)=\textrm{cos}\theta\Phi_{\eta_8}(x)+\textrm{sin}\theta\Phi_{\eta_0}(x),
\end{equation}
where $\theta$ is the mixing angle which is taken to be $19^\circ$. Considering $\Phi_{\eta_8}=(u\bar{u}+d\bar{d}-2s\bar{s})/\sqrt6$ and $\Phi_{\eta_0}=(u\bar{u}+d\bar{d}+s\bar{s})/\sqrt3$, the transition matrix element becomes
\begin{equation}
\begin{aligned}
       \langle D^0(P_1)\eta(P_2)|D^{*0}_2(P)\rangle
       &=(2\pi)^4\delta^{(4)}(P-P_1-P_2)(-\textrm{i}P^\mu_2) \\
       &\times(\frac{-2M^2_\eta\textrm{cos}\theta}{\sqrt6M^2_{\eta_8}f_{\eta_8}}+\frac{M^2_\eta\textrm{sin}\theta}{\sqrt3M^2_{\eta_0}f_{\eta_0}})\langle D^0(P_1)|\bar{s}\gamma_\mu\gamma_5s|D^{*0}_2(P)\rangle,
\end{aligned}
\end{equation}
where $f_{\eta_8}$ and $f_{\eta_0}$ are the decay constants of $\eta_8$ and $\eta_0$, respectively. And the mass of $\eta$ is related to the masses of $\eta_8$ and $\eta_0$ by $M^2_\eta=(\textrm{cos}^2\theta M^2_{\eta_8}-\textrm{sin}^2\theta M^2_{\eta_0})/(\textrm{cos}^4\theta-\textrm{sin}^4\theta)$. Hence the decay amplitude is
\begin{equation}
      T(D^{*0}_2\to D^0\eta)=(-\textrm{i}P^\mu_2)(\frac{-2M^2_\eta\textrm{cos}\theta}{\sqrt6M^2_{\eta_8}f_{\eta_8}}+\frac{M^2_\eta\textrm{sin}\theta}{\sqrt3M^2_{\eta_0}f_{\eta_0}})\langle D^0(P_1)|\bar{s}\gamma_\mu\gamma_5s|D^{*0}_2(P)\rangle.
\end{equation}

Within Mandelstam formalism \cite{Man} the hadronic transition matrix element can be written as~\cite{zhang, wzh01}:
\begin{equation}
\begin{aligned}
          \langle D^{(\ast)+}(P_1)&|d\gamma_\mu\gamma_5\bar{u}|D^{*0}_2(P)\rangle  =\int\frac{\textrm{d}\vec{q}}{(2\pi)^3}\textrm{Tr}\left[\overline{\varphi^{++}_{P_1}}(\vec{q}_1)\gamma_\mu\gamma_5\varphi^{++}_P(\vec{q})\frac{\slashed{P}}{M}\right],
\end{aligned}
\end{equation}
where $\vec{q}$ and $\vec{q}_1$ are the relative three-momenta between the quark and anti-quark in the initial and final mesons, respectively. They are related by $\vec{q_1}=\vec{q}-\alpha\vec{P_1}$ with $\vec{P_1}$ being the three-dimensional momentum of $D^{(\ast)+}$. $\varphi^{++}_P(\vec{q})$ and $\varphi^{++}_{P_1}(\vec{q}_1)$ are the positive-energy parts of the Salpeter wave functions of $D^{*0}_2$ and $D^{(\ast)+}$, respectively, whose definitions can be found in Refs.~\cite{kim,wgl09}. As for $\langle D^0(P_1)|\bar{s}\gamma_\mu\gamma_5s|D^{*0}_2(P)\rangle$, the formulae are the same.

After finishing the integral in Eq.~(8), we get the $D^{*0}_2\to D^{(\ast)+}\pi^-$ decay amplitude
\begin{equation}
\begin{aligned}
           T=\frac{-\textrm{i}}{f_{P_2}}\epsilon^{\mu\nu}P_{1\mu}P_{1\nu}t
\end{aligned}
\end{equation}
for the $2^+\rightarrow 0^-0^-$ channel and
\begin{equation}
\begin{aligned}
       T&=\frac{-\textrm{i}}{f_{P_2}}\epsilon^{\alpha\beta\gamma\delta}\epsilon_{\alpha\mu} \epsilon_{1\beta} P_\gamma P_{1\delta} P_1^\mu s
\end{aligned}
\end{equation}
for the $2^+\to1^-0^-$ channel, where $s$ and $t$ are corresponding form factors.

If the final light meson is $\eta$, then the decay amplitudes are
\begin{equation}
\begin{aligned}
           T=-\textrm{i}(\frac{-2M^2_\eta\textrm{cos}\theta}{\sqrt6M^2_{\eta_8}f_{\eta_8}}+\frac{M^2_\eta\textrm{sin}\theta}{\sqrt3M^2_{\eta_0}f_{\eta_0}})\epsilon^{\mu\nu}P_{1\mu}P_{1\nu}t'
\end{aligned}
\end{equation}
for the $2^+\rightarrow 0^-0^-$ channel and
\begin{equation}
\begin{aligned}
       T&=-\textrm{i}(\frac{-2M^2_\eta\textrm{cos}\theta}{\sqrt6M^2_{\eta_8}f_{\eta_8}}+\frac{M^2_\eta\textrm{sin}\theta}{\sqrt3M^2_{\eta_0}f_{\eta_0}}) \epsilon^{\alpha\beta\gamma\delta}\epsilon_{\alpha\mu} \epsilon_{1\beta} P_\gamma P_{1\delta} P_1^\mu s'
\end{aligned}
\end{equation}
for the $2^+\to1^-0^-$ channel. Here $s'$ and $t'$ are corresponding form factors as well.

By finishing two-body phase space integral, we get the decay widths
\begin{equation}
\begin{aligned}
& \Gamma_{D^{*0}_2\to D\pi}=\frac{|\vec{P_1}|}{8\pi M^2} \frac{1}{2J+1}\sum_\lambda|T(D^{*0}_2\to D\pi)|^2, \\
&\Gamma_{D^{*0}_2\to D^*\pi}=\frac{|\vec{P_1}|}{8\pi M^2} \frac{1}{2J+1}\sum_\lambda |T(D^{*0}_2\to D^*\pi)|^2,
\end{aligned}
\end{equation}
where $\lambda$ represents the summation of all the polarizations in the transition amplitude and $J$ is the spin quantum number of the initial meson, which is $2$ at this case. The formulae for decay $D^{*0}_2\to D^0\eta$ are the same.

\section{Results and discussion}

The wave functions of heavy-light mesons can be achieved by solving the instantaneous BS equation. As for instantaneous, we mean the time component of the relative momentum is set to 0. A Cornell-type interaction kernel is applied when solving the eigen equation numerically. Here we will not give the details of this equation which can be found in Ref.~\cite{kim}. We just present the values of constituent quark mass used in the calculation: $m_b=4.96$ GeV, $m_c=1.62$ GeV, $m_s=0.5$ GeV, $m_u=0.305$ GeV, and $m_d=0.311$ GeV. We take the meson mass to have the following values~\cite{pdg}: $M_\pi=0.139$ GeV, $  M_{K^+}=0.494$ GeV, $M_{K^0}=0.497$ GeV, $M(D^*_2(2460)^0)=2.463$ GeV, and $M(D^*_{s2}(2573)^+)=2.572$ GeV.  As to the decay constants of light pseudoscalar mesons, we use $f_\pi=0.1307$ GeV, $f_K=0.1561$ GeV~\cite{pdg}, $f_{\eta_8}=1.26f_\pi$, $f_{\eta_0}=1.07f_\pi$, $M_{\eta_8}=0.6047$ GeV, and $M_{\eta_0}=0.9230$ GeV~\cite{chang}.

Our results are presented in Table I, where other models' results are also listed in order to do comparison. Here $D^{(\ast)}\pi$  represents $D^{(\ast)+}\pi^0+D^{(\ast0)}\pi^+$, and it's the same for the $D^{(\ast)}K$ case. For $D_2^\ast(2460)^0$, our results are $\Gamma[D\pi]=34.8$ MeV and $\Gamma[D^\ast\pi]=16.4$ MeV, which are close to those of Ref.~\cite{rosner} (single-quark-transition fomalism), Ref.~\cite{god05} ($^3P_0$ model), Ref.~\cite{zz} (chiral quark model), and Ref.~\cite{cs} ($^3P_0$ model). Ref.~\cite{god16} also adopts the $^3P_0$ model, while different parameter values are used; Ref.~\cite{mat} uses the chiral quark model and relativistic wave functions for the heavy-light mesons; Ref.~\cite{dai} uses HQET and QCD sum rule. All the three references get decay widths smaller than ours. The $D\eta$ channel has the decay width about $0.1$ MeV, which is much smaller than those of the previous two channels. The E1 decay width give the main EM contributions for this state, which is a few tens of keV~\cite{cs, god05}. At this stage, it can be neglected completely. The total decay width can be approximated by the sum of the two dominant channels. Our result is close to the experimental value 47.4 MeV~\cite{pdg}.

For $D^\ast_{s2}(2573)^0$, we get the total decay width 19.5 MeV, which is close to the experimental value 16.9 MeV. As for the partial widths of two dominant channels $DK$ and $D^\ast K$, our results are close to those in Ref.~\cite{god05}, Ref.~\cite{seg} ($^3P_0$ model), and Ref.~\cite{zz}. Not like the $D^\ast_2(2460)^0$ case, Ref.~\cite{cs} gets a larger value than ours. Again the $D_s\eta$ and the E1 decay channels give much smaller contribution, which could be neglected when estimate the total decay width.

In Table II, we present the partial decay widths and branching ratios of different channels with specific charges. One can see the isospin symmetry is approximately fulfilled.
Experimentally, the ratio of two dominant decay widths is interesting. Our results are
\begin{equation}\begin{aligned}
R_1=\frac{\Gamma[D^\ast_2(2460)^0\to D^+\pi^-]}{\Gamma[D^\ast_2(2460)^0\to D^{\ast+}\pi^-]}= 2.13,
\end{aligned}\end{equation}
\begin{equation}\begin{aligned}
R_2=\frac{\Gamma[D^\ast_2(2460)^0\to D^+\pi^-]}{\Gamma[D^\ast_2(2460)^0\to D^{\ast+}\pi^-]+\Gamma[D^\ast_2(2460)^0\to D^+\pi^-]}= 0.68,
\end{aligned}\end{equation}
and
\begin{equation}\begin{aligned}
R_3=\frac{\Gamma[D^\ast_{s2}(2573)^+\to D^{\ast 0}K^+]}{\Gamma[D^\ast_{s2}(2573)^+\to D^{0}K^+]}= 0.08.
\end{aligned}\end{equation}
In PDG~\cite{pdg}, $R_1=1.54\pm 0.15$ is smaller than our result, and $R_2=0.62\pm 0.03\pm 0.02$ is consistent with ours. $R_3$ is much smaller than the experimental upper limit 0.33.

\begin{table}
\caption{The partial and total decay widths (MeV) of $D_2^\ast(2460)^0$ and $D_{s2}^\ast(2573)^+$, respectively. Ref.~\cite{zz} uses the chiral quark model. In Ref.~\cite{god16}, the mass values $M[D_2^\ast(1P)]=2502$ MeV and $M[D_{s2}^\ast(1P)]=2592$ MeV are used.}
\vspace{0.2cm}
\setlength{\tabcolsep}{0.01cm}
\centering
\begin{tabular*}{\textwidth}{@{}@{\extracolsep{\fill}}ccccccccc}
\hline\hline
\multirow{2}{*}{Modes}&\multicolumn{4}{c}{ $D^*_2(2460)^0$ }&\multicolumn{4}{c}{$D^*_{s2}(2573)^+$}\\
\cline{2-5}\cline{6-9}
&$D\pi$&$D^\ast\pi$&$D\eta$&$total$&$DK$&$D^\ast K$&$D_s\eta$&$total$
\\ \hline
{\phantom{\Large{l}}}\raisebox{+.2cm}{\phantom{\Large{j}}}
This work&34.8&16.4&0.0864&51.3&18.1&1.43&0.0667&19.6\\
{\phantom{\Large{l}}}\raisebox{+.2cm}{\phantom{\Large{j}}}
Ref.~\cite{rosner}&27.4&19.4&&46.8&&&&\\
{\phantom{\Large{l}}}\raisebox{+.2cm}{\phantom{\Large{j}}}
Ref.~\cite{mat}&10.4&4.0&&14.4&6.7&0.51&&7.2\\
{\phantom{\Large{l}}}\raisebox{+.2cm}{\phantom{\Large{j}}}
Ref.~\cite{god05}&37&18&&55&20&1&&21\\
{\phantom{\Large{l}}}\raisebox{+.2cm}{\phantom{\Large{j}}}
Ref.~\cite{seg}&&&&&16.71&1.88&0.08&18.67\\
{\phantom{\Large{l}}}\raisebox{+.2cm}{\phantom{\Large{j}}}
Ref.~\cite{zz}&39&19&0.1&59&16&1&0.4&17\\
{\phantom{\Large{l}}}\raisebox{+.2cm}{\phantom{\Large{j}}}
Ref.~\cite{cs}&35&20&0.08&55&27&3.1&0.2&30\\
{\phantom{\Large{l}}}\raisebox{+.2cm}{\phantom{\Large{j}}}
Ref.~\cite{god16}&15.3&6.98&0.107&23.0&9.40&0.545&0.105&10.07\\
{\phantom{\Large{l}}}\raisebox{+.2cm}{\phantom{\Large{j}}}
Ref.~\cite{dai}&13.7&6.1&&19.8&&&&\\
{\phantom{\Large{l}}}\raisebox{+.2cm}{\phantom{\Large{j}}}
PDG~\cite{pdg}&&&&$47.7\pm1.3$&&&&$16.9\pm0.8$\\
\hline\hline
\end{tabular*}
\end{table}

\begin{table}
\caption{Decay width and branching radio}
\vspace{0.2cm}
\setlength{\tabcolsep}{0.1cm}
\centering
\begin{tabular*}{\textwidth}{@{}@{\extracolsep{\fill}}ccc}
\hline\hline
Strong decay channel & Decay width (MeV) & Branching radio (\%)  \\
  \hline
  {\phantom{\Large{l}}}\raisebox{+.2cm}{\phantom{\Large{j}}}
        $D^*_2(2460)^0\rightarrow D^+\pi^-$ & 22.8 & 44.5 \\
        {\phantom{\Large{l}}}\raisebox{+.2cm}{\phantom{\Large{j}}}
        $D^*_2(2460)^0\to D^{*+}\pi^-$ & 10.7 & 21.0 \\
        {\phantom{\Large{l}}}\raisebox{+.2cm}{\phantom{\Large{j}}}
        $D^*_2(2460)^0\to D^{0}\pi^0$ & 12.0 & 23.4 \\
        {\phantom{\Large{l}}}\raisebox{+.2cm}{\phantom{\Large{j}}}
        $D^*_2(2460)^0\to D^{*0}\pi^0$ & 5.66 & 11.1 \\\hline
        {\phantom{\Large{l}}}\raisebox{+.2cm}{\phantom{\Large{j}}}
        $D^*_{s2}(2573)^+\to D^{0}K^+$ & 9.49 & 48.7  \\
        {\phantom{\Large{l}}}\raisebox{+.2cm}{\phantom{\Large{j}}}
        $D^*_{s2}(2573)^+\to D^{*0}K^+$ & 0.804 & 4.1  \\
        {\phantom{\Large{l}}}\raisebox{+.2cm}{\phantom{\Large{j}}}
        $D^*_{s2}(2573)^+\to D^{+}K^0$ & 8.61 & 44.1  \\
        {\phantom{\Large{l}}}\raisebox{+.2cm}{\phantom{\Large{j}}}
        $D^*_{s2}(2573)^+\to D^{*+}K^0$ & 0.630 & 3.2  \\
\hline\hline
\end{tabular*}
\end{table}

\section{conclusion}

In conclusion, we have calculated the two-body strong decay widths of $D^*_2(2460)^0$, and $D^*_{s2}(2573)^+$ by using PCAC and low energy theorem. The wave functions of the heavy-light mesons are achieved by solving corresponding instantaneous BS equations. The predicted total decay widths and the ratios of the partial widths of two dominant decay channels are close to the PDG values, except that $R_1$ is a little larger. The experimental results about the branching ratios are still missing and more data are expected to be accumulated.

\section*{Acknowledgments}
This work was supported in part by the National Natural Science
Foundation of China (NSFC) under Grant No.~11405037, No.~11575048, and No.~11505039.

\end{document}